\documentclass[aps, prl, a4paper, 10pt, showpacs, twocolumn]{revtex4-1}
\usepackage{amsmath, amssymb, amsthm, bm, bbm, textcomp}
\usepackage[T1]{fontenc}
\usepackage[latin9]{inputenc}
\usepackage{graphicx,graphics}
\usepackage{geometry}
\usepackage{color}
\newcommand{\one}{\mathbbm{1}}
\newcommand{\Tr}[1]{\mathrm{Tr}\left[ {#1} \right]}
\newcommand{\ket}[1]{\left|{#1}\right\rangle}
\newcommand{\bra}[1]{\left\langle{#1}\right|}

\newcommand{\be}{\begin{equation}}
\newcommand{\ee}{\end{equation}}
\newcommand{\eea}{\end{eqnarray}}
\newcommand{\bea}{\begin{eqnarray}}

\begin{document}

\title{Experimental violation of multipartite Bell inequalities with trapped ions}

\author{B. P. Lanyon$^{1,2}$\footnote{benban}, M. Zwerger$^3$, P. Jurcevic$^{1,2}$, C. Hempel$^{1,2}$, W. D\"ur$^3$, H. J. Briegel$^{1,3}$, R. Blatt$^{1,2}$, and C. F. Roos$^{1,2}$}
\affiliation{
$^1$ Institut f\"ur Quantenoptik und Quanteninformation der \"Osterreichischen Akademie der Wissenschaften, A-6020 Innsbruck, Austria\\
$^2$ Institut f\"ur Experimentalphysik, Universit\"at Innsbruck, Technikerstr. 25, A-6020 Innsbruck, Austria \\
$^3$ Institut f\"ur Theoretische Physik, Universit\"at Innsbruck, Technikerstr. 25, A-6020 Innsbruck,  Austria}
\date{\today}

\begin{abstract}

We report on the experimental violation of multipartite Bell inequalities by entangled states of trapped ions. First we consider resource states for measurement-based quantum computation of between 3 and 7 ions and show that all strongly violate a Bell-type inequality for graph states, where the criterion for violation is a sufficiently high fidelity. 
Second we analyze GHZ states of up to 14 ions generated in a previous experiment using stronger Mermin-Klyshko inequalities, and show that in this case the violation of local realism increases exponentially with system size. These experiments represent a violation of multipartite Bell-type inequalities of deterministically prepared entangled states. In addition, the detection loophole is closed.
\end{abstract}

%03.65.Ud Entanglement and quantum nonlocality (e.g. EPR paradox, Bell's inequalities, GHZ states, etc.)
%%03.67.Lx    Quantum computation
%03.67.Bg Entanglement production and manipulation
%37.10.Ty 	Ion trapping

%03.67.Mn Entanglement measures, witnesses, and other characterizations
%32.80.Qk 	Coherent control of atomic interactions with photons

\maketitle

% ---------------------------------------------------
% Introduction
% ---------------------------------------------------

\textit{Introduction ---}
How strong can physical correlations be? Bell inequalities set a bound on the possible strength of non-local correlations that could be explained by a theory based on some fundamental assumptions known as "local realism". Quantum mechanics predicts the existence of states which violate Bell's inequality, rendering a description of these states by a local hidden variable (LHV) model impossible. While first discovered for bipartite systems in a two-measurement setting \cite{Be64}, Bell inequalities have been extended to multi-measurement settings and multipartite systems, leading to a more profound violation for larger systems of different kinds \cite{We01,Mermin,Sc05,Gu05,Ca08}.

In particular, it was shown that all graph states violate local realism, where the possible violation increases exponentially with the number of qubits for certain types of states \cite{Gu05,Sc05,Ca08}. Graph states \cite{He04,He06} are a large class of multiqubit states that include a number of interesting, highly entangled states, such as the 2D cluster states \cite{Ra01b} or the GHZ states. They serve as resources for various tasks in quantum information processing, including measurement-based quantum computation (MBQC) \cite{Ra01,Br09} or quantum error correction \cite{CSS}. The results of \cite{Gu05,Sc05} (see also \cite{Di97}) provide an interesting connection between the usability of states for quantum information processing and the possibility to describe them by a LHV model.

Here we experimentally demonstrate the violation of multi-partite Bell-type inequalities for graph states generated with trapped ions. First we consider a range of graph states that find application in MBQC and observe strong violations in all cases. Second, for a different class of graph states, we investigate the scaling of the multi-partite Bell violation with system size and confirm an exponential increase: that is the quantum correlations in these systems become exponentially stronger than allowed by any LHV model. 

To be more precise, in the first part of our work we consider graph states that allow one to perform single-qubit and two-qubit gates in MBQC, as well as resource states for measurement-based quantum error correction \cite{La13}. That is, we demonstrate that not only the codewords of quantum error correction codes violate local realism \cite{Di97}, but also the resource states for encoding and decoding and other computational tasks. In this part we make use of general Bell-type inequalities derived for all graph states in Ref. \cite{Gu05}. We show that the Bell observable simply corresponds to the fidelity of the state, i.e. a violation is guaranteed by a sufficiently high fidelity. This allows the many previous experiments that quote fidelities to be reanalyzed to see if a Bell violation has been achieved.

For the purpose of investigating the scaling of Bell violations we consider a sub-class of graph states, for which stronger inequalities are available \cite{Mermin,Sc05,Ca08}, e.g. the Mermin-Klyshko inequalities for $N$-qubit GHZ states \cite{Mermin}. We show that these Mermin-Klyshko inequalities \cite{Mermin} are violated by GHZ states from 2 to 14 qubits generated in previous experiments \cite{Mo11}. In fact, we confirm an (exponentially) increasing violation with system size.

Multi-partite Bell violations for smaller system sizes were previously obtained with photons \cite{ExpPhotons}. Here specific 4-photon states encoding up to 6 qubits were considered. For trapped ions only two-qubit systems have previously been shown to violate a Bell inequality \cite{ExpIons}. Here we deal with larger systems and states with a clear operational meaning in measurement-based quantum information processing, where each qubit corresponds to a separate particle. Finally, our detection efficiency is such that we close the detection loophole.

\textit{Background ---}
\label{Sec_background}
Graph states $\ket{G}$ are defined via the underlying graph $G$, which is a set of vertices $V$ and edges $E$, that is $G=(V,E)$. One defines an operator $K_j= X_j \prod_{i \in N(j)}Z_i$ for very vertex $j$, where $X$ and $Z$ denote Pauli spin-$\frac{1}{2}$ operators. $N(j)$ denotes the neighborhood of vertex $j$ and is given by all vertices connected to vertex $j$ by an edge. The graph state $\ket{G}$ is the unique quantum state which fulfills $K_j\ket{G} = \ket{G}$ for all $j$, i.e. it is the common $+1$ eigenstate of all operators $K_j$. An equivalent definition starts with associating a qubit in state $\ket{+}=1/\sqrt{2}\left(\ket{0}+\ket{1}\right)$ with every vertex and applying a controlled phase (CZ) gate between every vertices connected by an edge, $\ket{G}={\cal U}_G|+\rangle^{\otimes n}$ with ${\cal U}_G=\prod_{(k,l) \in E} CZ^{(k,l)}$.
Graph states have important applications in the context of measurement-based quantum computation as resource states \cite{Ra01,Br09} and quantum error correction \cite{CSS}.

In \cite{Gu05} it was shown that all graph states give rise to a Bell inequality and that the graph state saturates it. Thus neither the correlations nor the quantum information processing that exploits these correlations can be accounted for by a LHV model. The inequality is constructed in the following way. One aims at writing down an operator ${\cal{B}}$ (specifying certain correlations in the system) such that the expectation value for all LHV models is bounded by some value ${\cal{D}}$, while certain quantum states yield an expectation value larger than ${\cal D}$. Let $S(G)$ denote the stabilizer \cite{Go96} of a graph state $\ket{G}$. It is the group of the products of the operators $K_j$ and is given by $S(G)= \{s_j , j=1,..., 2^n \}$ with $s_j=\prod_{i \in I_j(G)}K_i$ where $I_j(G)$ denotes a subset of the vertices of $G$. For the state corresponding to the empty graph, the generators of the stabilizer group are given by $K_j=X_j$, and the stabilizer group is given by all possible combinations of $X$ and $\one$ on the different qubits. For $n=2$ we have $S(G) =\{\one\otimes\one, X\otimes\one, \one\otimes X, X\otimes X \}$. Notice that for any non-trivial graph states (i.e. graph states with a non-empty edge set $E$), these operators are simply transformed via ${\cal U}_G K_j {\cal U}_G^\dagger$ since $|G\rangle = {\cal U}_G|+\rangle^{\otimes n}$, where $U_G X_j U_G^\dagger = X_j \prod_{i \in N(j)}Z_i$, i.e. the stabilizing operators of the graph state specified above.

\begin{figure}[ht]
\centering
\vspace{5mm}
\includegraphics[scale=0.35]{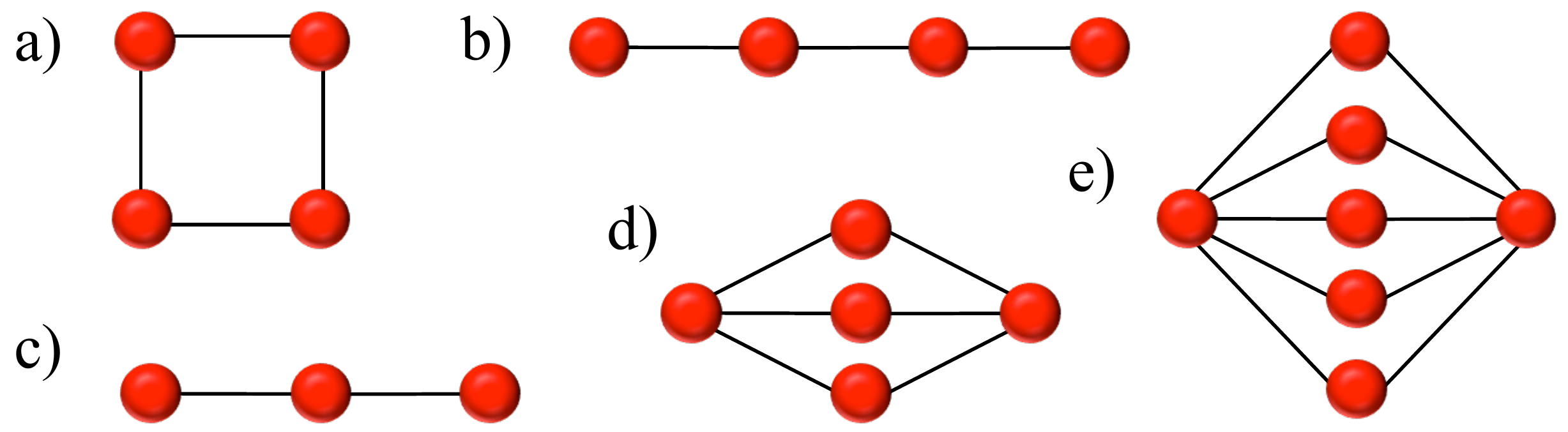}
\vspace{0mm}
\caption{Graph states that find application in measurement-based quantum computation. Red circles represent qubits, connecting lines relate to the states' generation method, as described in the text. a) Box graph $\ket{BC_4}$ b) Linear graph $\ket{LC_4}$. Error Correction graphs:  c) $\ket{EC_1}$ d) $\ket{EC_3}$ e) $\ket{EC_5}$}
\label{FigGraphs}
\vspace{0mm}
\end{figure}

The normalized Bell operator is defined as ${\cal{B}}_n(G)=\frac{1}{2^n} \sum_{i=1}^{2^n}s_i(G)$, and we have $\langle {\cal{B}}_n(G) \rangle_\rho \leq 1$ (where, in quantum mechanics, $\langle {\cal{B}}_n(G) \rangle_\rho=Tr[{\cal{B}}_n(G)\rho]$ for density matrix $\rho$). 
Let ${\cal{D}}(G)=\operatorname{max}_{\operatorname{LHV}}|\langle{\cal{B}}_n\rangle|$ where the maximum is taken over all LHV models. For any non-trivial graph state ${\cal{D}}(G)<1$ \cite{Gu05}. 
The maximization is generally hard to perform, but has been explicitly carried out for graph states with small $n$ in \cite{Gu05}. The basic idea is to assign a fixed value ("hidden variable") $+1$ or $-1$ to each Pauli operator $X_j,Y_j,Z_j$, and determine (numerically) the setting that yields a maximum value of ${\cal B}_n(G)$. This then also provides an upper bound on all LHV models. The corresponding Bell inequality reads 
\be
\langle {\cal{B}}_n(G) \rangle \leq {\cal D}(G),
\label{bell}
\ee
which is non-trivial whenever ${\cal{D}}(G) < 1$. For the states $\ket{LC_4},\ket{BC_4},\ket{EC_1}$ one finds ${\cal{D}}=0.75$ \cite{Gu05}, while we show in \cite{Sup} that ${\cal D}(EC_3) \leq 0.75$ and ${\cal D}(EC_5) \leq 0.625$ (see figure \ref{FigGraphs} for the different states). For fully connected graphs corresponding (up to a local basis change) to $n$-qubit GHZ states $|{\rm GHZ}_n\rangle =(|0\rangle^{\otimes n} + |1\rangle^{\otimes n})/\sqrt{2}$, we obtain ${\cal D}({\rm GHZ}_n) = 1/2 + 2^{-n/2}$ for $n \leq 14$ (see \cite{Sup}).

Any graph state $\ket{G}$ fulfills $\bra{G}{\cal{B}}_n(G)\ket{G}=1$, since the state is a $+1$ eigenstate of all operators appearing in the sum that specifies ${\cal B}_n(G)$. Hence it follows that the graph state maximally violates the graph Bell inequality (\ref{bell}), $\bra{G}{\cal{B}}_n(G)\ket{G} > {\cal{D}}(G)$. 

A straightforward calculation shows that the normalized Bell operator equals the projector onto the graph state: ${\cal{B}}_n(G) = \frac{1}{2^n} \sum_{i=1}^{2^n} s_i = \ket{G}\bra{G}$. This can be seen directly for the empty graph by noting that $|+\rangle \langle +|= (\one + X)/2$, and writing out the product for $|+\rangle \langle +|^{\otimes n} = \prod_{j=1}^n (\one_j + X_j)/2$ which yields all combinations of $X$ and $\one$. The result for a general graph state follows by transforming each operator via ${\cal U}_G X_j {\cal U}_G^\dagger=K_j^G$, together with $|G\rangle ={\cal U}_G|+\rangle^{\otimes n}$. Thus, the expectation value $\langle {\cal{B}}_n(G) \rangle$ equals the fidelity $F(\rho_{G_{exp}}) = \operatorname{Tr}(\rho_{G_{exp}}\ket{G}\bra{G})$, where $\rho_{G_{exp}}$ denotes the density matrix of the experimentally obtained graph state. As it is common practice to report on the fidelity this provides a simple way of reinvestigating earlier experiments.

In addition, this provides a possibility for measuring the fidelity of a graph state by measuring the $2^n$ stabilizers, which add up to ${\cal{B}}_n$. Although this method has the same exponential scaling behavior as full state tomography, it requires significantly fewer measurement settings.

\textit{Results: Graph states for MBQC ---}
The first group of graph states that we consider are resources for MBQC and are shown in figure \ref{FigGraphs}. The four-qubit box cluster $\ket{BC_4}$ represents the smallest element of the 2D cluster (family) required to implement arbitrary quantum algorithms \cite{Ra01b,Ra01,Br09}. The four-qubit linear cluster state $\ket{LC_4}$ can be used to demonstrate a universal quantum logic gate set for MBQC \cite{La13, Wa05}. The graph states $\ket{EC_n}$ allow for the demonstration of an $n$-qubit measurement-based quantum error correction code \cite{La13}.

Except for $\ket{BC_4}$, all of these states were generated in a system of trapped ions and their application to MBQC was demonstrated in our recent paper \cite{La13}. In that work, and in particular its accompanying supplementary material, one can find information on the experimental techniques used to prepare the states. In summary, $n$ qubits are encoded into the electronic state of $n$ $^{40}$Ca$^{+}$ ions held in a radio-frequency linear Paul trap: each ion represents one qubit. After preparing each qubit into the electronic and motional ground state, graph states are generated deterministically and on demand using laser pulses which apply qubit-state dependent forces to the ion string. Additional details relevant to Bell inequality measurements are now described. The ions are typically 6 $\mu$m apart and it takes approximately 500 $\mu$s to generate the states. Individual qubits can be measured in any basis with near unit fidelity in 5 $m$s. The state $\ket{BC_4}$ belongs to the same family as the error correction graphs, i.e.  $\ket{BC_4}=\ket{EC_2}$, and was thus generated using exactly the method described in \cite{La13}.

For each $n$-qubit graph state shown in figure \ref{FigGraphs} we experimentally estimate each of the $2^n$ expectation values $\langle s_i(G) \rangle$ that are required to estimate $\langle {\cal{B}}_n(G) \rangle$. If this final number is larger than allowed by LHV models then the multi-partite Bell inequality is violated. The experimental uncertainty in each $\langle s_i(G) \rangle$ is the standard quantum projection noise that arises from using a finite number of repeated measurements to estimate an expectation value.

We note that the full density matrices for a subset of the graph states shown in figure \ref{FigGraphs} were presented in \cite{La13}. We do not extract the data from these matrices but directly measure the $2^n$ observables in each case. No previous characterization of the states $\ket{BC_4}$ and $\ket{EC_5}$ has been done.

The results are summarized in table \ref{stats} and clearly show that all experimentally generated states violate their graph state inequalities by many tens of standard deviations. Recall that $\langle {\cal{B}}_n(G) \rangle$ is equal to the state fidelity. For comparison, table \ref{stats} also presents the state fidelity measured in another way --- by reconstructing the density matrix $\rho_{exp}$ via full quantum state tomography and using $Tr(\ket{G}\bra{G}\rho_{exp})$. This approach is much more measurement-intensive, requiring the estimation of $3^n$ expectation values and was therefore not carried out for the 7-qubit state $\ket{EC_5}$. The fidelities derived in these different ways are seen to overlap to within 1 standard deviation.  In the supplementary material we give an explicit example of how the experimental value of $\langle {\cal{B}}_n(G)\rangle$ for one graph state ($\ket{BC_{4}}$) was derived.

\begin{table*}[ht]
%\begin{table*}[htbd]
\caption{
\label{tableB} \textbf{Properties of experimentally generated graph states}.
Fidelity $F{=}\Tr{[\rho\ket{\psi}\bra{\psi}]}$ derived from the tomographically reconstructed density state ($\rho$), where $\ket{\psi}$ is the ideal state. $\langle{\cal{B}}_n \rangle$ is equivalent to the state fidelity, derived from a subset of tomographic measurements. Values on the rhs of the inequality are the maximum allowed by LHV models (${\cal{D}}(G)$, see \cite{Sup}). ${\cal{R}}={\cal{B}}_n/{\cal{D}}$ denotes the relative violation of the Bell inequality. NM: not measured. Errors are one standard deviation and derived from quantum projection noise.
}
\begin{center}
\begin{tabular}{|c|c|c|c|c|c|}
\hline
Graph & qubits & Fidelity ($F$)  & Multipartite Bell inequality $\langle {\cal{B}}_n \rangle$ & relative violation ${\cal{R}}$ \\
\hline\hline
$LC_4$  & 4 & 0.841$\pm{0.006}$  & 0.85$\pm{0.02}> 0.75$ & $1.13\pm0.03$  \\
$BC_4$ & 4 & 0.847$\pm{0.007}$ & 0.86$\pm{0.02}> 0.75$ & $1.15\pm0.03$  \\
$EC_1$ & 3 & 0.920$\pm{0.005}$  & 0.92$\pm{0.02}> 0.75$ & $1.23\pm0.03$  \\
$EC_3$ & 5 & 0.843$\pm{0.005}$  & 0.86$\pm{0.01}> 0.75$ & $\geq 1.15\pm0.01$  \\
$EC_5$ & 7 & NM  & 0.73$\pm{0.01}>0.625$ & $\geq1.17\pm0.02$\\
\hline
\end{tabular}
\end{center}
\label{stats}
\end{table*}%

\begin{figure}[t]
\centering
\includegraphics[scale=0.5]{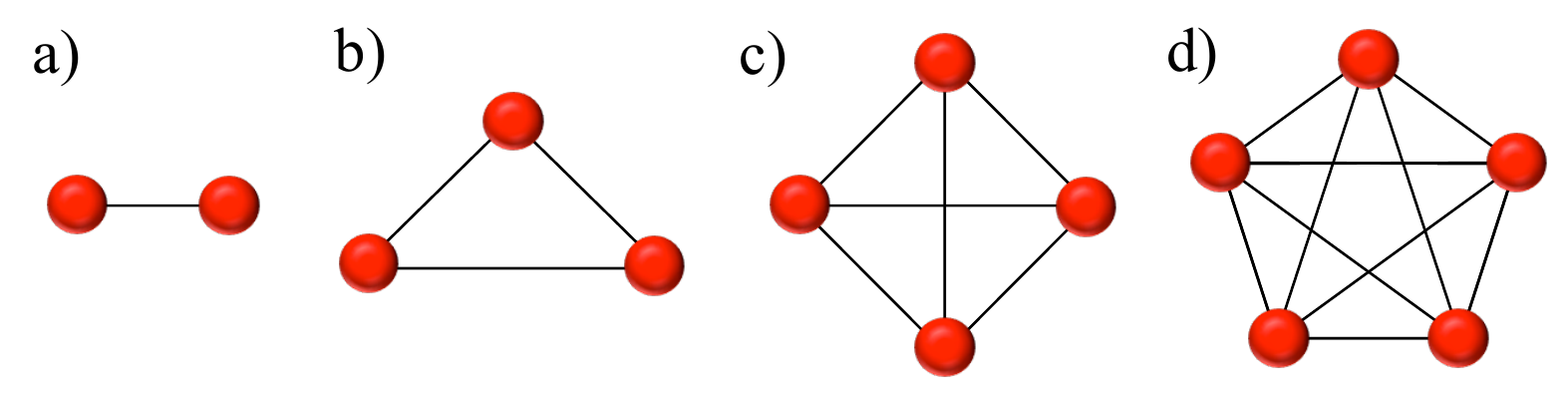}
\caption{Examples of (fully connected) graph states that are (up to local unitary operations) equivalent to $n$-qubit GHZ states with $n=2,\ldots ,5$. Red circles represent qubits, connecting lines relate to the states' generation method, as described in the text.
}
%\vspace{-5mm}
\label{GHZ}
\end{figure}

\textit{Results: scaling of violation with system size ---}
In the second part of our work we are interested in investigating the scaling of the violation of multi-partite Bell inequalities with the system size. Table \ref{stats} presents the relative violation observed for the graph state inequalities, defined as the ratio of the quantum mechanical expectation value of the Bell observables and the maximal reachable value in a LHV model (${\cal{R}}= \langle {\cal{B}}_n(G) \rangle / {\cal{D}}(G)$). From this it is clear that while all the generated MBQC graph states violate their inequalities, the size of the violation does not change significantly with the size of the graph state. However, there is another class of Bell inequalities, the Mermin-Klyschko (MK) inequalities \cite{Mermin}, for which the quantum mechanical violation is predicted to increase exponentially with qubit number. The MK inequalities apply to the GHZ states $|{\rm GHZ}_n\rangle =(|0\rangle^{\otimes n} + |1\rangle^{\otimes n})/\sqrt{2}$, which are (up to local unitary operations) equivalent to graph states corresponding to a fully connected graph (see figure \ref{GHZ}).

\begin{figure}[t]
\vspace{0mm}
\includegraphics[scale=0.4]{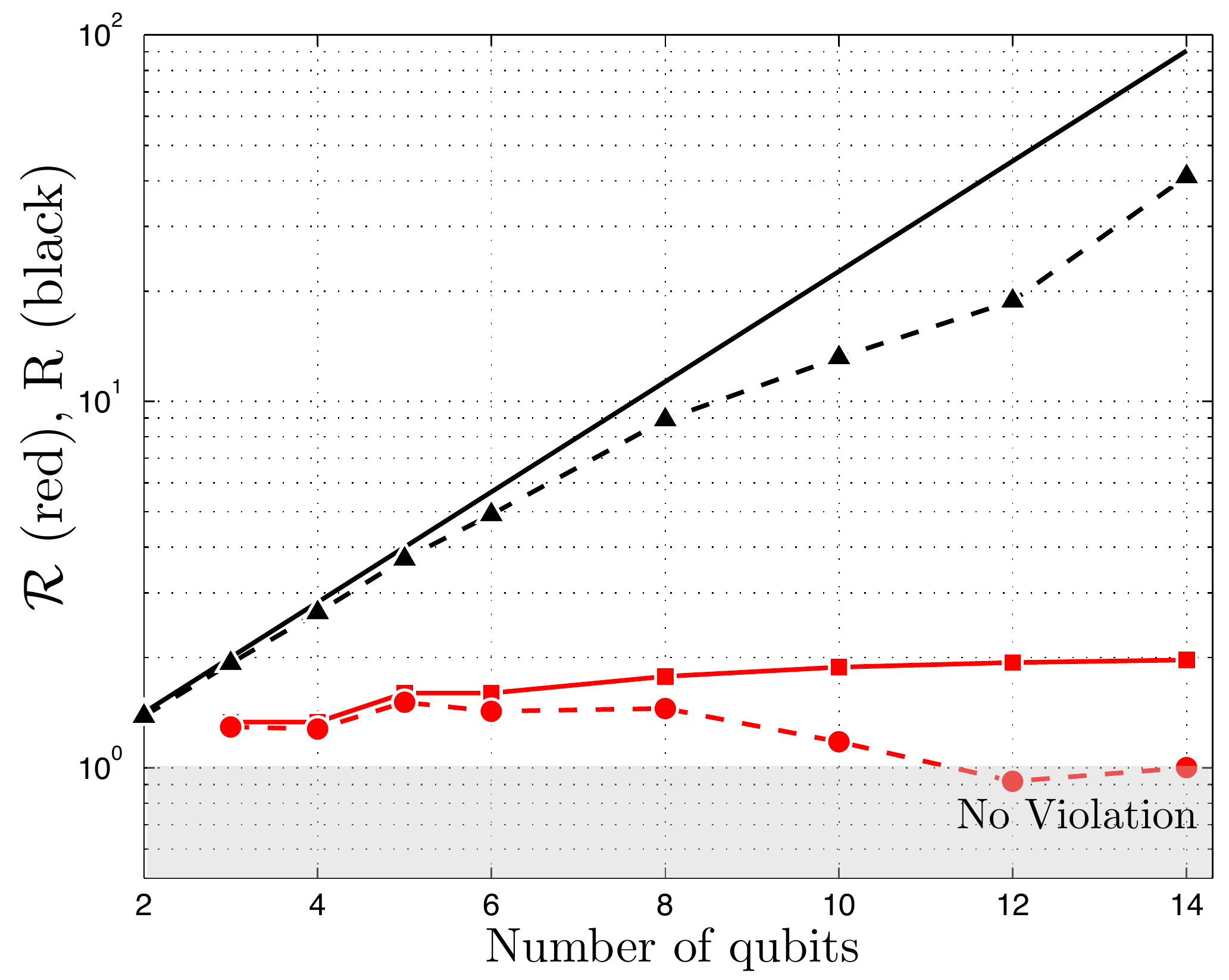}
\vspace{0mm}
\caption{Multipartite Bell inequality violations for GHZ states of different sizes. Data is taken from Ref. \cite{Mo11}.
$\cal{R}$ is the relative violation via the ratio of the quantum mechanical expectation value of the Bell observables and the maximal reachable value in a LHV model.
It is given by ${\cal{R}}= \langle {\cal{B}}_n(G) \rangle / {\cal{D}(G)}$ for the graph inequalities (red lines)  and $R=  \langle B_n \rangle / D = 2^{(n-1)/2} \langle B_n \rangle$ for the MK inequalities (black lines).
In each case: solid (dashed) lines show the ideal (experimental) case.
Error bars in experimental results are all smaller than the point sizes.
Any value larger than ${\cal{R}}{=}1$ corresponds to a Bell violation. Notice the logarithmic scaling of the axis.
 }
\label{scaling}
\end{figure}

The MK Bell operator \cite{Mermin} can be defined recursively by
\be
B_k=\frac{1}{2\sqrt{2}} B_{k-1} \otimes (\sigma_{a_k}+\sigma_{a'_k}) + \frac{1}{2\sqrt{2}} B'_{k-1} \otimes (\sigma_{a_k}-\sigma_{a'_k})
\ee
and starts with $B_1=\sigma_{a_1}$ \cite{footnoteNorm}.
The $\sigma_{a_k}$ are given by scalar products of three dimensional unit vectors $\bold{a_k}$ and the vector $\bm{\sigma}$ consisting of the three Pauli operators, i.e. $\sigma_{a_k}=\bold{a_k} \cdot \bm{\sigma}$. The operator $ B'_k$ is obtained from $B_k$ by exchanging all the $a_k$ and $a'_k$. Within a LHV model one can only reach $D=\operatorname{max}_{\operatorname{LHV}}|\langle B_n\rangle| = 2^{-(n-1)/2}$ \cite{Mermin}. This can be seen intuitively by assigning specific values $+1$ or $-1$ to each of the operators $\sigma_{a_k}, \sigma_{a'_k}$, which implies that the recursive relation reduces to $B_k = \pm\frac{1}{\sqrt{2}} B_{k-1}$ or $B_k = \pm \frac{1}{\sqrt{2}} B'_{k-1}$ where $B_1 = \sigma_{a_1} = \pm 1$ for all possible choices. It follows that $D= 2^{-(n-1)/2}$ in this case, and similarly for all LHV models.

Quantum mechanics allows a violation of the MK inequality by $\langle B_n \rangle = 1$; by comparison to the maximum allowed LHV value $D$, one sees that the violation scales exponentially with the system size. 
Note that the MK inequality achieves the highest violation for any inequality with two observables per qubit \cite{We01}. The observables can be significantly simplified by choosing the same measurement directions for all qubits, e.g. $\sigma_{a_j}=X$ and $\sigma_{a_j^{'}}=Y$ for all $j$. It can then be shown that \cite{Mermin}
\be
B_n=(e^{i\beta_n}|1\rangle^{\otimes n}\langle 0|+e^{-i\beta_n}|0\rangle^{\otimes n}\langle 1|),
\ee
with $\beta_n\equiv (n-1)\pi/4$. The determination of $\langle B_n \rangle$ then reduces to determining two specific off-diagonal elements in the density matrix $\rho$.
The states which violate the MK inequality maximally are then given by $\ket{\psi_n} = 1/\sqrt{2} ( \ket{0}^{\otimes n} + e^{i\beta_n} \ket{1}^{\otimes n} )$, leading to $\bra{\psi_n} B_n \ket{\psi_n}=1$. Notice that the local observables can be adjusted in such a way that GHZ states with arbitrary phase $\beta_n$ maximally violate the corresponding MK-inequality, i.e. the relevant quantity for a violation is given by the absolute value of the coherences $|0\rangle^{\otimes n}\langle 1|$.

GHZ states of the form $\ket{\psi_n}$ for up to $n=14$ qubits have previously been prepared using trapped ions \cite{Mo11} (again 1 qubit is encoded per ion). In that work the state fidelities were estimated via measurements of the logical populations $\ket{0}^{\otimes n}\bra{0}$ and $\ket{1}^{\otimes n}\bra{1}$, and the coherences $|0\rangle^{\otimes n}\langle 1|$. From this information both the graph state Bell observable $\langle {\cal B}_n(G) \rangle$ and the MK Bell observable $\langle B_n \rangle$ can now be calculated.

The relative violations $R$, defined as $R=  \langle B_n \rangle / D = 2^{(n-1)/2} \langle B_n \rangle$ for the MK inequalities and ${\cal R}=  \langle {\cal B}_n(G) \rangle / {\cal D}(G)$ for the graph inequality, are presented graphically in figure \ref{scaling}. An exponential scaling is apparent for the relative violation $R$ of the MK inequalities, i.e. by using larger systems a stronger violation of non-locality can be observed. We now show that the violation of the MK inequalities with larger systems can be more robust to noise than for smaller systems. This can be illustrated as follows. Assume the preparation of a noisy $n$-qubit GHZ state, where imperfections and decoherence is modeled in such a way that each qubit is effected by single qubit depolarizing noise ${\cal E}_j(p)\rho = p \rho + (1-p)/4 \sum_{k=0}^{3} \sigma_k^{(j)} \rho \sigma_k^{(j)}$, i.e. $\rho = \prod_{j=1}^n {\cal E}_j |{\rm GHZ_n}\rangle \langle {\rm GHZ_n}|$. Even though the state can be  shown straightforwardly to have an exponentially small fidelity, one nevertheless encounters a violation of the MK inequality even for a large amount of local depolarizing noise. To be specific, one finds that ${\rm tr}(B_n \rho)= p^n$ (the off-diagonal elements are simply suppressed by this factor), leading to $R=(\sqrt 2 p)^n/\sqrt{2}$. That is, as long as $p > 1/{\sqrt{2}}$, one encounters a violation of the MK inequality for large enough $n$. This means that MK inequalities can tolerate almost $30\%$ noise per qubit. The graph inequalities for GHZ states demand a fidelity larger than 0.5 \cite{Sup}, requiring the noise per qubit to reduce exponentially with system size.

\textit{Conclusion and outlook ---}
We have demonstrated the violation of multi-partite Bell inequalities for graph states which are resources in MBQC, thereby confirming a connection between applicability of states as resources for quantum information processing and violation of LHV models. In addition, we show that the data in a previous experiment is sufficient to identify an exponentially increasing Bell violation with system size \cite{Mo11}. Given the fact that our set-up can readily be scaled up to a larger number of ions, this opens the possibility to demonstrate LHV violations for large-scale systems.

\textit{Acknowledgements ---}
This work was supported by the Austrian Science Fund (FWF): P25354-N20, P24273-N16 and SFB F40-FoQus F4012-N16.

%\newpage
%\section*{}
%\newpage%\newpage
%\section*{}
%\newpage

\section{appendix}
\section{Examples of Bell operators (theory)}

For illustration, we explicitly provide some of the Bell operators, both for graph state inequalities and MK inequalities. As an example, for the graph $LC_4$, i.e. the linear cluster state of four qubits, the normalized Bell operator ${\cal{B}}_4(LC_4)$ is given by

\begin{eqnarray}
{\cal{B}}_4(LC_4) & = & \frac{1}{16} (\mathbb{I}\mathbb{I}\mathbb{I}\mathbb{I} + XZ\mathbb{I}\mathbb{I} + ZXZ\mathbb{I} + \mathbb{I}ZXZ \nonumber \\
 &  & {} + \mathbb{I}\mathbb{I}ZX + YYZ\mathbb{I} + X\mathbb{I}XZ + XZZX \nonumber \\
 &  & {} + ZYYZ + ZX\mathbb{I}X + \mathbb{I}ZYY - ZYXY \nonumber \\
 &  & {} + X\mathbb{I}YY + YY\mathbb{I}X - YXYZ + YXXY).\nonumber \\
\end{eqnarray}

For the graph $BC_4$, i.e. the box cluster state of four qubits, the normalized Bell operator ${\cal{B}}_4(BC_4)$ is given by

\begin{eqnarray}
{\cal{B}}_4(BC_4) & = & \frac{1}{16} (\mathbb{I}\mathbb{I}\mathbb{I}\mathbb{I} + XZ\mathbb{Z}\mathbb{I} + ZX\mathbb{I}Z + 
Z\mathbb{I}XZ \nonumber \\
 &  & {} + \mathbb{I}ZZX + YYZZ + YZYZ + X\mathbb{I}\mathbb{I}X \nonumber \\
 &  & {} + \mathbb{I}XX\mathbb{I} + ZYZY + ZZYY - \mathbb{I}YYX \nonumber \\
 &  & {} - Y\mathbb{I}XY - YX\mathbb{I}Y - XYY\mathbb{I} + XXXX).\nonumber \\
 \label{bc4}
\end{eqnarray}

For the four-qubit GHZ state and the corresponding MK inequalities the Bell operator $B_4$ is given by   

\bea
B_4 & = & -\frac{1}{8} (YXXY + YXYX + YYXX - YYYY \nonumber \\
& & + XYYX + XYXY + XXYY - XXXX ).
\eea

\section{A complete experimental example}

\begin{figure}[htb]
\centering
\includegraphics[scale=0.15]{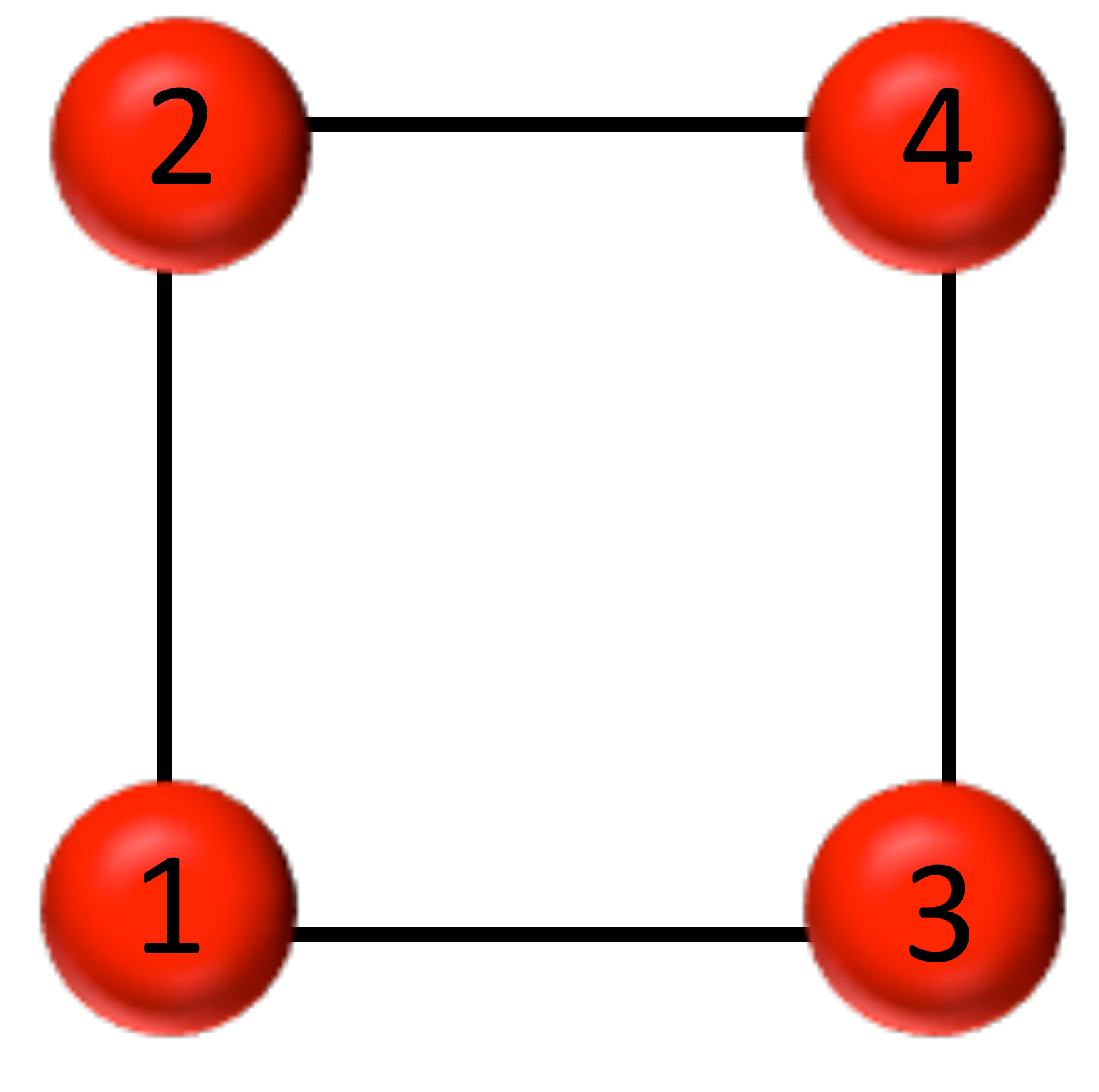}
\caption{Four-qubit box cluster $\ket{BC4}$, showing the qubit number labeling employed.}
\label{box}
\end{figure}

In this section we provide details on how the Multipartite Bell inequality $\langle {\cal{B}}_n \rangle$, given in the table in the main text, was measured for one of the graph states. Specifically we choose the 4-qubit box cluster $\ket{BC4}$ shown in figure \ref{box}.  As described in the main text, the most well known method to prepare clusters states is to initialize each physical qubit in the state $\ket{+}{=}(\ket{0}+\ket{1})/\sqrt{2}$ and then to apply CP gates between every pair of qubits with a connecting edge: in this case qubit pairs 1\&2, 1\&3, 2\&4 and 3\&4. Note that

\begin{equation}
CP=e^{-iH_{cp}\frac{\pi}{4}},
\end{equation}

where $H_{cp}{=}(\mathbb{I}-Z)\otimes(\mathbb{I}-Z)$.  

In our experiments we prepare all our graph states in a different way, which is equivalent to the method using CP gates up to single qubit rotations: i.e. the states are equivalent up to a local change of basis. In summary we begin by initializing all qubits into $\ket{1}$ and applying pairwise entangling operations generated by Hamiltonians of the form $H_{ms}=(X)\otimes(X)$, where the subscript $ms$ refers to the M{\o}lmer-S{\o}rensen interaction on which our qubit interactions are based \cite{MS99}. 
For more experimental details on the state generation see the supplementary material of \cite{La13} where laser pulse sequences can be found. Note that the 4-qubit box cluster is not presented in \cite{La13}, however the laser pulse sequence is identical to that for all the error-correction states $\ket{EC_n}$. In fact $\ket{BC4}=\ket{EC_2}$: rotating the box cluster diagram in figure \ref{box} by 45 degrees in either direction (so that it becomes a diamond) makes it clear that it belongs to the same family of states. 

As stated, experimentally we do not prepare $\ket{BC4}$, but ideally a locally rotated state for which we will use the label $\ket{\hat{BC4}}$. This state is given by

\begin{equation}
\ket{\hat{BC4}}=\frac{\ket{0000}-\ket{0110}-\ket{1001}-\ket{1111}}{2}
\end{equation}

which is equivalent to the state $\ket{BC4}$ made with CP gate once it is corrected by the following single-qubit correction rotations. qubit 1: HXZ, qubit 2: HX, qubit 3: HX, qubit 4 HXZ, where H is the Hadamard, and X and Z are standard Pauli operators.

For the experimentally generated 4-qubit box cluster $\ket{\hat{BC4}}$, the normalized Bell operator ${\cal{B}}_4(\hat{BC_4})$ is given by

\begin{eqnarray}
{\cal{B}}_4(\hat{BC_4}) & = & \frac{1}{16} (\mathbb{I}\mathbb{I}\mathbb{I}\mathbb{I} + IYYZ -\mathbb{I}XXZ 
+\mathbb{I} ZZ\mathbb{I} \nonumber \\
 &  & {} +Y\mathbb{I} ZY + YYXX + YXYX + YZ\mathbb{I} Y \nonumber \\
 &  & {} -X\mathbb{I} ZX + XYXY + XXYY +XZ\mathbb{I} X\nonumber \\
 &  & {}  +Z\mathbb{I} \mathbb{I} Z+ZYY\mathbb{I} -ZXX\mathbb{I} X+ZZZZ).\nonumber \\
 \label{bc4}
\end{eqnarray}

The experimentally observed expectation values for all 16 observables are presented in table \ref{boxresults}. The average values of the last column is $0.86\pm{0.02}$ and is the normalized Bell operator we observe for this state. All uncertainties are one standard deviation and derive from the intrinsic uncertainty in using a finite number of measurements to estimate expectation values.

\begin{table*}[ht]
%\begin{table*}[htbd]
\caption{
\label{tableB} \textbf{Results for 4-qubit ring cluster}.
Experimentally we prepare the state $\ket{\psi}=(\ket{0000}-\ket{0110}-\ket{1001}-\ket{1111})/2$, which is equivalent to the 4-qubit ring cluster made with CPHASE gates once it is corrected by the following single-qubit correction operators. qubit 1: HXZ, qubit 2: HX, qubit 3: HX, qubit 4 HXZ, where H is the Hadamard, and X and Z are standard Pauli operators. The following 16 observables were measured and the outcomes for the ideal and measured cases are presented.
}
\begin{center}
\begin{tabular}{ |c|c|c|c|}
  \hline 
 Number& Operator $\hat{O}$& Trace($\hat{O}\rho_{ideal}$)&Trace($\hat{O}\rho_{exp})$\\
  \hline
   1& Z    Z    Z    Z    &1.000    &0.8700$\pm{0.0626}$\\
   2&- Z    X    X    $\mathbb{I}$   &1.000   &0.8700$\pm{0.0628}$\\
    3&Z    Y    Y    $\mathbb{I}$    &1.000   & 0.8500$\pm{0.0623}$\\
    4&Z    $\mathbb{I}$    $\mathbb{I}$    Z    &1.000    &0.9100$\pm{0.0626}$\\
    5&X    Z    $\mathbb{I}$    X   &1.000  & 0.8300$\pm{0.0625}$\\
    6&X    X    Y    Y    &1.000    &0.8100$\pm{0.0668}$\\
    7&X    Y    X    Y    &1.000    &0.7800$\pm{0.0670}$\\
    8&-X    $\mathbb{I}$    Z    X   &1.000   &0.8500$\pm{0.0625}$\\
    9&Y    Z    $\mathbb{I}$    Y   & 1.000   & 0.8700$\pm{0.0629}$\\
    10&Y    X    Y    X    &1.000    &0.7300$\pm{0.0671}$\\
    11&Y    Y    X    X    &1.000   & 0.8800    $\pm{0.0665}$\\
    12&Y    $\mathbb{I}$    Z    Y    &1.000    &0.8300    $\pm{0.0629}$\\
    13&$\mathbb{I}$    Z    Z    $\mathbb{I}$    &1.000    &0.8800    $\pm{0.0626}$\\
    14&-$\mathbb{I}$    X    X    Z   &1.000   &0.8600    $\pm{0.0628}$\\
   15& $\mathbb{I}$    Y    Y    Z    &1.000   & 0.8600    $\pm{0.0623}$\\
   16& $\mathbb{I}$    $\mathbb{I}$    $\mathbb{I}$    $\mathbb{I}$    &1.000    &1.00$\pm{0.0626}$\\
   \hline  
\end{tabular}
\end{center}
\label{boxresults}
\end{table*}%

\section{Values for ${\cal{D}}(G)$ for $\ket{EC_n}$}

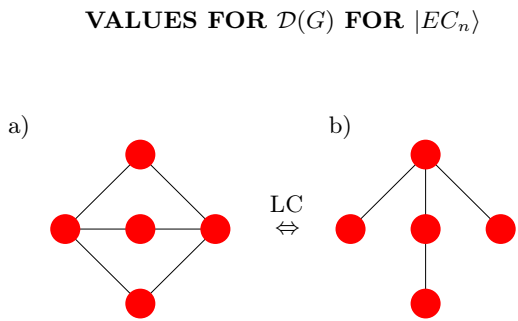
\begin{figure}[htb]
\centering
\includegraphics[scale=0.35]{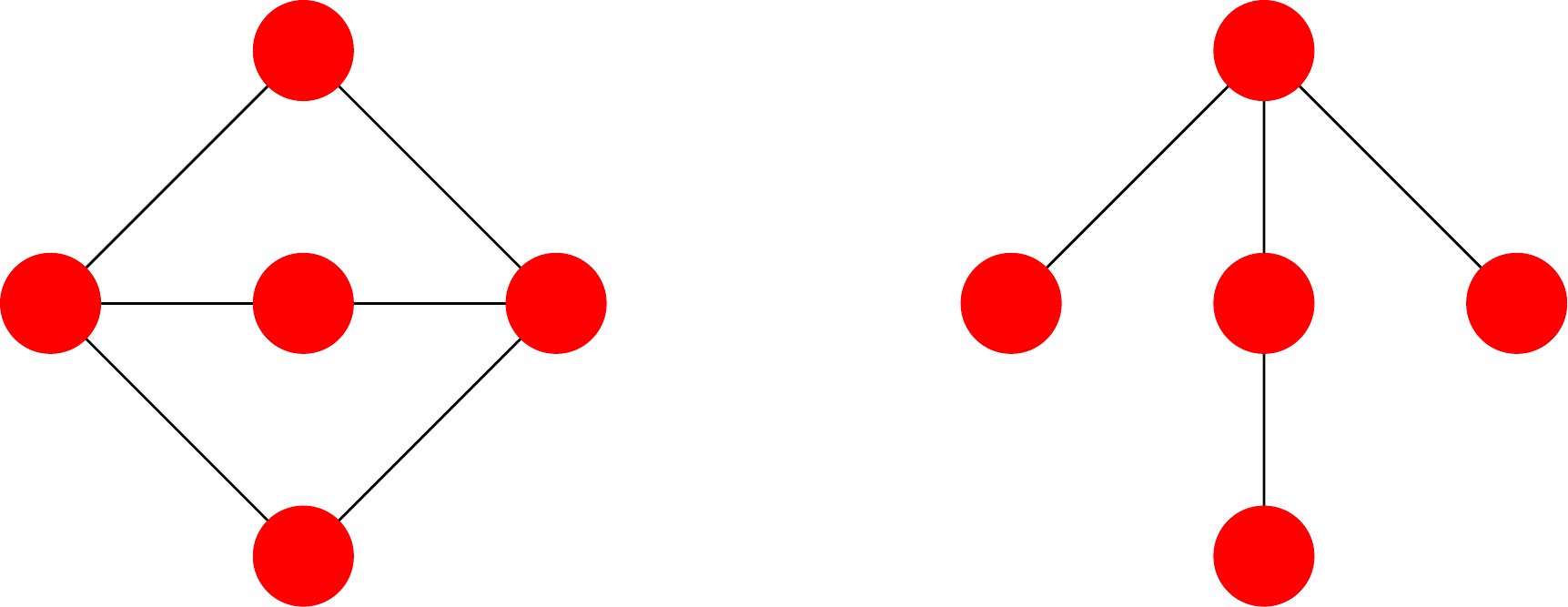}
\put(-190,70){a)}
\put(-70,70){b)}
\put(-90,32){$\Leftrightarrow$}
\put(-92,40){LC}
\caption{The two graph states are LC equivalent, that is they differ only by local Clifford operations. a) $\ket{EC_3}$ b) $\ket{EC_{3LC}}$.}
\label{bell5}
\end{figure}

A bound for ${\cal{D}}(EC_3)$, where $EC_3$ is the graph underlying the five qubit state $\ket{EC_3}$ which we used to demonstrate quantum error correction, can be found in the following way. First one notes that $\ket{EC_3}$ is equivalent to the state $\ket{EC_{3LC}}$ in figure \ref{bell5}b) up to local Clifford (LC) operations. The two graph states have the same rank indices and are thus equivalent up to local unitary operations \cite{He04}. The fact that they are both graph states then implies the LC equivalence. The local Clifford operations do not change the value of ${\cal{D}}(EC_3)$. The graph state $\ket{EC_{3LC}}$ is build from a four qubit GHZ state $\ket{GHZ_4}$ and a single qubit graph $\ket{G_1}$, connected by an edge. Application of Lemma 3 in \cite{Gu05} then gives a bound on ${\cal{D}}(EC_3)$:

\begin{equation}
{\cal{D}}(EC_3) \leq {\cal{D}}(G_1) {\cal{D}}(GHZ_4) = \frac{3}{4}.
\end{equation}

In a similar way one can bound the value ${\cal{D}}(EC_5)$,

\begin{equation}
{\cal{D}}(EC_5) \leq {\cal{D}}(G_1) {\cal{D}}(GHZ_6) = \frac{5}{8}.
\end{equation}

\section{Values for ${\cal{D}}(G)$ for GHZ states}

The values for ${\cal{D}}(G)$ for GHZ states with up to ten qubits have been derived numerically in \cite{Gu05}. Here we illustrate how one can simplify the numerical procedure and provide the values for GHZ states with twelve and fourteen qubits. In addition we show that a fidelity larger than one half is required for all GHZ states in order to violate the graph state inequality.

The generators for GHZ states are given (up to irrelevant local Clifford operations) by $K_1=XZZ \ldots Z$, $K_2=ZXZ \ldots Z$, \ldots, $K_n=ZZ \ldots ZX$. The Bell operator contains all products of the generators, as described in the main text. In \cite{Gu05} it is shown that one can restrict to LHV models which assign $+1$ to all $Z$ measurements. For GHZ states one can then show by simply multiplying the generators that one only has to check the following operators: for stabilizers with an odd number $j_{\rm{odd}}$ of generators: 

\begin{equation}
O_{\rm{odd}}=(-1)^{(j_{\rm{odd}}-1)/2} X^{\otimes j_{\rm{odd}}} \otimes \mathbb{I}^{\otimes n-j_{\rm{odd}}},
\end{equation} 
and for stabilizers with an even number $j_{even}$ of generators:
\begin{equation}
O_{\rm{even}}=Y^{\otimes j_{\rm{even}}} \otimes \mathbb{I}^{\otimes n-j_{\rm{even}}},
\end{equation}
and all the permutations of the qubits in both cases.

Since each of the operators $O_{\rm{odd}}$ and $O_{\rm{even}}$ contain only $X$ or $Y$ operators, they can be optimized independently. For the operators $O_{\rm{even}}$ it is easy to see that they contribute maximally by assigning $+1$ to all $Y$ measurement outcomes. Their total contribution to ${\cal{D}}$ is then given by $2^{-n}\sum_{k=0}^{n/2} \tbinom{n}{2k} = 1/2$, where the factor $2^{-n}$ comes from the normalization in the definition of ${\cal{B}}_n$ and the sum comes from the total number of operators $O_{\rm{even}}$. The optimization for the operators $O_{\rm{odd}}$ is done numerically and we find ${\cal{D}}(GHZ_{12})=33/64$ and ${\cal{D}}(GHZ_{14})=65/128$. For even $n \leq14$ one can confirm ${\cal{D}}(GHZ_n)= 1/2 + 2^{-n/2}$. We leave it as a conjecture that this expression holds for arbitrary even $n$.

The contribution from the operators $O_{even}$ puts a lower bound on ${\cal{D}}$ and thus, via the relation $\langle {\cal{B}}_n \rangle_{\rho_{exp}} = F(\rho_{exp})$, on the fidelity $F$. Consequently, a necessary requirement for any GHZ state to violate the Bell type inequality derived in \cite{Gu05}  is that the fidelity is greater than one half.


\begin{thebibliography}{99}
\bibitem{footnoteNorm}
Notice that we have included a normalization factor such that $\langle B_k \rangle_\rho \leq 1$.
\end{thebibliography}
\end{document}